\documentstyle[emulateapj]{article}  %  ApJ journal style

\lefthead{Fabbiano et al.}
\righthead{{\it Chandra} ACIS Observations of NGC 4038/4039}

\def\ergcm2s{erg cm$^{-2}$ s$^{-1}$} % ergs per cm**2 second
\def\etal{et al.}		% Present ApJ style, override older definitions
\def\n4038{NGC 4038/39}		% Use in text.  For title, use NGC 4038/4039
\def\chandra{{\it Chandra}}
\begin{document}

\title{Chandra Observations of ``The Antennae'' Galaxies (\n4038)}

\author{G. Fabbiano, A. Zezas, and S. S. Murray }
\affil{Harvard-Smithsonian Center for Astrophysics,\\
60 Garden Street, Cambridge, MA 02138}

\bigskip

\begin{abstract}
We report the results of a deep \chandra\ ACIS pointing at the
merging system \n4038. We detect an extraordinarily luminous 
population of X-ray sources,  with luminosity well above that of
XRBs in M31 and the Milky Way. If these sources are unbeamed XRBs,
our observations may point to them being 10-(a few)100~$M_{\odot}$ black hole counterparts.
We detect an X-ray bright hot ISM, with features including
bright superbubbles associated with the actively star-forming knots, 
rehions where hot and warm (H$\alpha$) ISM intermingle, and
a large-scale outflow.
\end{abstract}


\section{Introduction}

%Hierarchical galaxy formation relies on galaxy mergers so the
%properties of low-$z$ mergers can constrain galaxy formation at high
%$z$.  They can supply an empirical connection between N-body
%simulations and the star formation history of the Universe.  
The Antennae (modeled by Toomre \& Toomre 1972 and Barnes
1988) are the nearest pair of colliding galaxies
involved in a major merger ($D$~= 29 Mpc, for $H_0$~= 50).  Hence,
this system provides a unique opportunity for getting the most
detailed look possible at the consequences of a galaxy merger as
evidenced by induced star formation and the conditions in the ISM.

Each of the two colliding disks shows rings of giant H~II regions and bright
stellar knots with luminosities up to $M_V \sim -16$ (Rubin et al 1970),
which are resolved with  the  {\it Hubble Space Telescope}
into typically about a dozen young star clusters
(Whitmore \& Schweizer 1995). These knots coincide with the peaks of
H$\alpha$, 2.2$\mu$, and 6-cm radio-continuum emission  
(Amram et al 1992; Stanford et al 1990; Hummel \& van der Hulst 1986),
indicating an intensity of star formation exceeding
that observed in 30 Doradus.  CO aperture synthesis maps reveal major
concentrations of molecular gas, including $\sim$2.7$\times 10^9
M_{\odot}$ in the region where the two disks overlap
(Stanford et al 1990; Wilson et al 2000). 
A recent K-band study derives ages for the star clusters ranging from 4 
to 13~Myrs, and measures high values of extinction ($A_V \sim 0.7 - 4.3$~mag;
Mengel et al 2000).

X-ray emission from the Antennae was first observed with the {\it
Einstein} IPC (Fabbiano et al 1982; Fabbiano \&
Trinchieri 1983), with a luminosity of $L_X \sim 1 \times
10^{41} \rm ergs~s^{-1}$, larger than the average X-ray luminosity of normal 
spiral galaxies (e.g. Fabbiano 1989 and refs. therein) .  
More recently the Antennae have been
studied with the ROSAT PSPC (Read et al 1995), ROSAT
HRI (Fabbiano et al 1997), and ASCA (Sansom et al 1996).
These instruments reveal different emission components, including soft
gaseous emission, which Read et al suggested may be connected with 
a bipolar outflow, and
harder emission. The 5'' resolution ROSAT HRI image reveals
very complex and intricate X-ray emission, clearly dominated by
components associated with the recent star formation activity,
and suggests both an exceptional population of super-luminous 
point-like sources (3 were detected with 
$L_X \sim 6 \times 10^{39} - 1 \times 10^{40} \rm ergs~s^{-1} $), 
and a possibly extended and filamentary hot ISM (Fabbiano et al 1997).  

In this paper we present the first results of a deep \chandra\ ACIS
(Garmire 1997, Weisskopf et al 2000) observation of the Antennae. The  spatial resolution of \chandra\ is 
10 times superior to that of the ROSAT HRI, allowing us to resolve
the emission on physical sizes of $\sim 70$~pc, versus the $\sim 700$~pc
resolution possible with the ROSAT HRI. With this resolution we can 
easily detect individual X-ray sources and image in detail 
the spatial properties
of more extended emission regions. At the same time, ACIS
allows us to study the X-ray spectral properties of these sources and
emission regions, providing additional important constraints on their nature,
something that was not possible with the HRI.

Here we summarize the overall results, as they pertain to the
general morphology of the X-ray emission and to the detection of
the different compact and gaseous X-ray emitting components.
A detailed analysis of the point source component 
of \n4038\ will be presented in  Zezas et al (2001, Paper~II).
A detailed  spectral analysis of the diffuse component 
and a multi-wavelength comparison of the ISM will be presented
in Fabbiano et al (2001, Paper~III).

\section{Observations and Data Analysis}

\chandra\ was pointed to \n4038\ on December 1, 1999, for 72.5~ks with
the back-illuminated ACIS-S3 CCD chip at the focus (Observation ID: 315). 
ACIS was at a temperature of -110 C during these observations.
The satellite telemetry
was processed at the \chandra\ X-ray Center (CXC) 
with the Standard Data Processing (SDP) system,
to correct for the
motion of the satellite and to apply instrument calibration. 
The data 
used for this work were processed in custom mode with the version R4CU5UPD6.5
of the SDP, to take advantage of improvements in processing software
and calibration, in advance of data reprocessing.  Verification 
of the data products showed no anomalies.
The data products were then analyzed with the CXC
CIAO software. CIAO Data Model tools were
used for data manipulation, such as screening out bad pixels, and producing
images in given energy bands.
The April 2000 release of the ACIS CCD calibration files (FEF) was used
for the analysis.

In preparation for further analysis the data were screened
to exclude the two `hot' columns present in ACIS-S3 at chip-x columns 512 
and 513 (at the b and c node boundaries). 
Besides causing stripes in the image, these hot columns contaminate the source
spectrum at the low energies (near 0.2~keV), as can be seen from fig.~1
where we show the integrated source count spectral distribution before
and after the screening. 

No screening to remove high background data
was necessary: the light curve extracted from an area of 7~arcmin$^2$ showed a constant 
count rate of $0.456\pm 0.004$ cts~s$^{-1}$, consistent with a low-level 
background radiation.

After completion of this work, the reprocessed Chandra data products became available.
While these data are generally consistent with those used in this paper, and 
therefore our results and conclusions stand, the aspect calibration was improved giving 
an improved absolute source position with 1$\sigma$ error of 0''.6.
The R4CU5UPD6.5 data instead have an absolute position error of $\sim 1''.8$, which
we have estimated by using stars in the field detected in X-rays. The revised
absolute positions are consistent with our astrometry.
We have used the latest and best Chandra positional information for our comparison with
other positional information (optical, CO, ROSAT HRI, H$\alpha$) later on in this paper.

\subsection{Characteristics of the Overall X-ray Emission}

Fig.~2 shows the full-band (0.1-10~keV) ACIS-S3 field including the Antennae galaxies.
The data here are shown without any processing, beside that described 
above. It is clear that the emission is complex, consisting of diffuse
emission regions of varying intensity and sizes, and of a number of bright
point-like sources. On fig.~2 are displayed a number of regions. The polygon,
surrounding the emission area associated with the Antennae galaxies, was 
chosen to be inclusive of the area covered by the optical images of these
galaxies; this region was used to extract the total source counts
and spectral counts as discussed later in this paper. The circles surrounding
this polygon are the background extraction areas for these estimates.
Point-like sources were excluded from the background areas (dashed circles).

Fig.~3a displays an adaptively smoothed image of the full-band 
ACIS image of \n4038,
generated with the CIAO task {\it csmooth},  which adjusts
the smoothing kernel in order to preserve a constant S/N across the
image. Each pixel in the input image is smoothed on its "natural" scale, 
in the sense that the smoothing scale is increased until the total number of 
counts under the kernel exceeds a value that is determined from a preset 
significance and the expected number of background counts in the kernel area.  
In our case, {\it csmooth} was set to perform the background calculation locally.
The smoothing function we used is a variable-width two-dimensional 
gaussian and the S/N was set to range between 3 and 5 over the image. 
In fig.~3b we also show an outline of the 
optical image of the interacting galaxy disks (white contours), 
the positions of the two nuclei (`X'), the positions of peaks
of the CO distribution, indicating the areas of highest obscuration
(crosses; Stanford et al 1990), and circles representing the source extraction areas
of the three possibly
point-like sources detected with the ROSAT HRI (X-3, X-11, X-12; Fabbiano et al 1997).
Comparison with the ROSAT PSPC, ASCA, and Einstein images is meaningless
given the significantly lower resolution of these telescopes 
(see Fabbiano et al 1997 for a comparison between ROSAT HRI and
earlier, lower angular resolution data). 
In general, however, our results confirm early findings pointing
to a complex X-ray emission field.
The X-ray image shows clearly a number of
bright point-like sources,  extended emission associated with the numerous 
star-formation regions in the two disks and with the nuclear regions, and 
lower surface brightness emission associated with the entire stellar system,
and possibly extending beyond it.

Fig.~4 is a true X-ray color image, where red corresponds to 0.1-2.0~keV,
green to 2.0-5.0~keV and blue to 5.0-10.0~keV. 
Images in each of the three bands
were adaptively smoothed, as described above. To produce this figure, the three images
were then combined using the CIAO tool {\it dmimg2jpg}, which
makes a color JPEG image (or EPS) from three image files.  
This color image shows that the bright
point-like sources have harder spectra than the diffuse emission: they appear white, 
implying that they emit in all three energy bands, with the exception of a 
`blue' heavily absorbed source, in the area of most
intense CO emission (Stanford et al 1990; see fig.~3b, also note some `green'
emission nearby, also indicative of X-ray absorption). The diffuse emission
instead is softer (red in the image), but it also shows some harder areas, for
example in the vicinity of the nucleus of NGC~4039 (the southern galaxy; see fig.~3b
for the position of the nucleus).

Fig.~5 shows the spectral distribution of the detected counts from \n4038 (black), 
the coadded  spectrum of detected sources (blue), which is dominated by the
luminous point-like sources, but includes a few faint, possibly extended
sources (see below), a representative bright point-like 
source spectrum (green),
and the spectrum of the diffuse emission (red), obtained by excluding the 
contributions of point-like and small-scale extended sources from the image.
The area used for the extraction of the integrated galaxian
emission is shown by the polygon in fig.~6 (also fig.~2).
The background counts were derived from the circular regions
marked in fig.~2.
The detected source regions that were subtracted from
the overall source area are shown by circles and ellipses in fig.~6.
The four largest ellipses (dashed in fig.~6) indicate
the presence of diffuse
and possibly complex emission, and were excluded from the spectral count extraction.

A detailed discussion of source detection and point-source spectral 
analysis is given in Paper II.
We find that there is a range of source spectra, but the most luminous
ones tend to be quite hard. This is consistent with what is shown in fig.~5,
where the point-source subtraction results in the lack of emission
from the higher energy channels.
The spectrum of the diffuse emission has features pointing to the 
optically thin thermal emission of a hot plasma 
(Fe-L complex; Mg~K$\alpha$; Si~K$\alpha$).
Note that in fig.~5 the integrated galaxian spectrum becomes noise-dominated
around 5~keV, while the coadded and individual bright source
spectra are still significant at higher energies. This is because the general
integrated spectrum is derived from a large area, that becomes background
dominated at the higher energies. The individual detected-source counts instead 
were extracted from the small source areas of fig.~6. 

\subsection{Point-like Sources}

We detect 48 sources at the 3$\sigma$ level with CIAO 
{\it wavdetect}, using scales from 1 to 16 pixels (0.5-8''); 
43 are detected in the 0.1-10.0~keV band, 2 only in the 2.0-4.0~keV and
and 3 only in the 0.1-2.0~keV band; 10 of these 48 sources
can be identified with small-size  features of the extended emission regions, while
the rest appear point-like. The latter 
include all the bright sources visible in the images we have shown 
earlier in this paper.  
Point-like detections range from 10 to 2063 counts per source, 
and account for
35.6\% of the entire \n4038\ counts in the 0.1-10.0~keV band. In the
`soft' (0.1-2.0~keV) band point sources produce 31.5\% of the total
emission whereas they dominate the X-ray emission in the `hard' band 
(2.0-10.0keV), accounting for 81.4\% of the total.  At a distance of 29~Mpc,
the `point-like' individual sources have an intrinsic 
luminosity ranging from $1.3\times10^{38} \rm erg~s^{-1}$ 
to $2.0\times10^{40} \rm erg~s^{-1}$, where we have used a 5~keV
bremsstrahlung model with galactic neutral hydrogen column
density along the line of sight ($N_H = 3.4\times10^{20}\rm cm^{-2}$)
for all the sources detected in the 0.1-10.0~keV band, and spectra
suggested by the spectral analysis (Paper~II) for the soft and mid-range
detected sources. Since these sources
are likely to be absorbed because of the dusty environment in the host 
galaxies (most remarkably, the `blue' source in fig.~4), or may have 
some intrinsic absorption, these
luminosities are likely to be lower limits to the real emitted source
luminosities. Full details of the source detection analysis and results are
given in Paper~II.

Three point-like sources were reported by Fabbiano et al
(1997) from the analysis of the ROSAT HRI image. Of these (see fig.~3b), source X-3
is coincident with a \chandra\ source; X-11 includes both a bright 
\chandra\ source and diffuse emission; X-12 includes 2 \chandra\ sources.
Note however that the ROSAT positions may be affected by a few arcseconds
aspect uncertainties, and also by the fact that the ROSAT HRI
sources were detected with 20-27 counts, which also include sizeable 
contributions from whatever else is in the source region. Under these
circumstances the source centroid would not be very accurate, because of
detection statistics, and would also be biased by the surrounding emission.

The great majority of the detected point-like sources are 
associated with \n4038.  Using the \chandra\ Deep Field (Giacconi et al 2000)
estimates, the number of background sources expected from a comparable area 
($\sim~5.9$~square arcminutes) is 1.7 at our 
soft band ($< 2$keV) flux threshold of $4 \times 10^{-16} \rm ergs~cm^{-2}~s^{-1}$. 
In this band we detect 34 point-like sources.
At our hard band ($> 2$keV) flux 
threshold of $2.8 \times 10^{-15} \rm ergs~cm^{-2}~s^{-1}$, 
we expect 1.4 background sources, and we detect 20 sources in the Antennae.
Given the presence of extended emission local to 
NGC~4038/39, we estimate that our source detection completeness limit is
a few $10^{38}\rm ergs~s^{-1}$.

\subsection{Extended X-ray Emitting Regions}

The {\it Chandra} images show extended emission features with sizes
ranging from a few arcseconds (few hundred parsecs) to scales of the
order of the entire galaxian system or larger. A complete study of
the spatial and spectral characteristics of this emission requires
a better calibration of the low energy response function of {\it Chandra} ACIS
than was available at the time of the present work, and so will
be deferred to Paper III. We can still, however, summarize the 
principal characteristics of this emission.

As mentioned in $\S$2.2, 10 of the relatively small-size sources detected
in the Antennae are not point-like. Some of these sources are easily identifiable
as peaks of the diffuse emission component 
within the Antennae `heads'.  These extended bright emission regions 
have luminosities in the $10^{39-40} \rm ergs~s^{-1}$ range,
assuming a thermal Raymond-Smith spectrum of kT=0.8~keV and 
galactic line of sight $N_H$.

The two nuclear regions are among the extended sources. 
At least in the case of the southern nucleus, the emission temperature
 may be higher that that of the emission associated with 
star forming knots in the arms (see fig.~4, the X-ray color map, that shows some
`white' in the southern nucleus area). Assuming a thermal spectrum with kT=5~keV and
line of sight $N_H$, we obtain luminosities of 1.3$ \times 10^{40} \rm ergs~s^{-1}$
and 1.5$ \times 10^{40} \rm ergs~s^{-1}$ for the 
northern and southern nuclear regions respectively. The nuclear regions may be 40\%
fainter if the emission is significantly softer (kT$\sim 0.8~keV$), as it is
more likely to be the case for the northern nucleus.

We examined the ACIS image in the soft band (0.1-2.0~keV)
for evidence of the large scale galaxian outflow (or halo)
suggested by Read et al (1995) on the basis of the low
angular resolution ($\sim$25'') ROSAT PSPC image.
While our results are very preliminary and need to  be confirmed 
with a proper spatial/spectral analysis of the extended soft
emission (Paper III), we note some low surface brightness, but
significant, emission extending mostly to the South of the
optical body in the ACIS image. This is shown in fig.~7, where we display the results of
an adaptive smoothing of a wide field section of the ACIS-S3 chip, 
with the data previously binned by a factor of 2, resulting in 1'' image pixels. 
Analysis of the raw data shows that the X-ray surface brightness excesses seen in these
regions range from 40\% to 10\% over the lower surface brightness (background)
areas in the image. These excesses cannot be explained with instrumental
non-uniformities, because the maximum {\it expected} excursions
from similar areas are of 5\% at the most, based on the instrumental
Quantum Efficiency Uniformity map and mirror vignetting.
In the South, for the same spectral assumptions as those used for
the soft diffuse emission within the Antennae, we estimate a luminosity for this 
very extended low-surface-brightness component
of $\sim (7.5 \pm 0.2) \times 10^{39} \rm ergs~s^{-1}$, assuming a Raymond-Smith 
spectrum with kT=0.8~keV and $N_H=3.4 \times 10^{20} \rm cm^{-2}$. 

\subsection{Luminosities of the Emission Components}

With the same spectral assumptions as above for the different galaxian 
components, the total (0.1-10)~keV luminosity of \n4038\ 
is 2.3$\times 10^{41} \rm ergs~s^{-1}$. The total point-source contribution is
1.1$\times 10^{41} \rm ergs~s^{-1}$, roughly half of the luminosity,
and the thermal extended components (including the nuclear regions) 
account for the remaining 1.2$\times 10^{41} \rm ergs~s^{-1}$.
Considering the amount of guesswork involved in earlier estimates, that
did not have the benefit of \chandra's resolution, these estimates are
in general agreement with those of Fabbiano et al (1997).

\section{Discussion}

The \chandra\ observation of the Antennae demonstrates how exceptional 
the X-ray emission of merging galaxies may be, in comparison with that 
of individual relatively undisturbed spiral systems (e.g. Fabbiano 1989, 1995). 
In the Antennae we detect a
population of super-luminous discrete X-ray sources, a very
luminous hot ISM, and two bright complex starburst nuclear regions.
While these components were suggested by the ROSAT HRI observations (Fabbiano et al 1997), it is
only with the combined subarcsecond spatial and spectral resolution of \chandra\
ACIS that we can firmly establish their presence and properties.

\subsection{A population of massive Black Hole Binaries?}

Fig.~8 shows the luminosity distribution of `point-like' sources detected in the
ACIS observation, compared with the upper end of the luminosity distribution 
of X-ray sources (most likely accretion binary systems, XRBs) in M31 (from {\it
Einstein}, Trinchieri
\& Fabbiano 1991; note that given the distances of the two systems, the 
physical resolution of the \chandra\ observation of \n4038\ is within a 
factor of 2 that of the {\it Einstein} HRI of M31).
In the case of \n4038\ we plot two histograms, coresponding to two different
choices of $H_o$: the filled one for $H_o = 50$, and the dashed one for $H_o = 75$.
The M31 and the \n4038\ histograms have been arbitrarily plotted in different scales, for ease
of display.  The point of this comparison is not
to compare the total number of fainter sources in the two systems, since
the different incompleteness biases prevent it. However, in both cases,
there are no such effects at luminosities higher than a few $10^{38} \rm ergs~s^{-1}$.

In M31, and the Milky Way (e.g. Watson 1990), XRBs have luminosities 
consistent with the Eddington limit of $\sim 1$ solar mass accreting object
($\sim 1.3 \times 10^{38} \rm ergs~s^{-1}$).
In the Antennae instead we find 14-8 (depending on $H_o$) sources with 
luminosities of $1 \times
10^{39} \rm ergs~s^{-1}$ or above, reaching as high as $\sim 10^{40} \rm ergs~s^{-1}$.
One of these sources (X-3) was detected, as possibly variable, with ROSAT (Fabbiano et al 1997).

Super-luminous `Super-Eddington' sources were detected in some nearby galaxies
prior to \chandra\ (e.g. Fabbiano 1995), and for some of them at least there is spectral
evidence  of accretion disks from ASCA, pointing to a black-hole binary counterpart (Makishima et al
2000). These sources tended to be associated with star-forming regions, but they
were rare, and the lack of angular resolution allowed the possibility of bright
clumps of X-ray sources in most cases.

What we see now is a population of these sources in a single galaxian system. As
discussed by Fabbiano et al (1997), their large X-ray luminosities exclude that we
are seeing the integrated stellar coronal emission of the star-forming knots.
These luminous sources have hard spectra, reminiscent of galactic
XRBs, very unlike the thermal spectra of the more diffuse emission 
regions (figs.~4, 5; Paper~II). 
Given their luminosities, that are comparable with, or exceed, the total 
integrated luminosity of M31, it is unlikely that they could be  clumps of
`normal' XRBs. We would need a few tens to a hundred neutron star binaries
(the entire X-ray emitting population of a normal spiral galaxy), clustered
in areas with 100~pc typical sizes. 

If these sources are single XRBs, their luminosities suggest 10-100 (or larger) $M_{\odot}$
black-hole counterparts, unless they are all beamed. 
The detection of soft excesses in their spectra (as
in the sources studied by Makishima et al 2000) would favor the
black-hole hypothesis. This will have to await the soft-band calibration of ACIS. 
It is possible that at least some of these 
super-luminous sources may be very young Supernovae (e.g. Schlegel 1995). Future 
monitoring of the Antennae is needed to verify this possibility. Any detection
of variability, other than secular fading, would point to XRBs, e.g.
as in the galaxy Holmberg-II, where the compact object is estimated to be a 
$\sim200$M$\odot$ black-hole candidate (Zezas \etal, 1999), or as recently
reported for M81 X-9 (La~Parola et al 2001).
Similarly luminous sources, albeit in smaller number, are detected in M82 with \chandra.
In M82, the detection of variability (Kaaret et al 2000,  Zezas et al 2000, Matsumoto 
et al 2001) demonstrates their XRB nature.

The presence of a copious number of these superluminous sources in galaxies 
undergoing violent starbursts, such as \n4038\ and M82, and their absence from
more settled galaxies, with an older stellar population, suggests that they are
short-lived objects. This may be consistent with a fast-evolving 
 massive black hole binary
hypothesis. This scenario has an interesting corollary: if 
what we see in the Antennae today was the norm at the epoch of galaxy formation,
then any spiral galaxy, such as our own Milky Way, may have an invisible
population of massive black holes, the remnants of primordial massive stars 
and of spent super-luminous XRBs. A number of these systems may have sunk
to the nucleus (Tremaine et al 1975) and contributed to the formation 
of massive nuclear black holes.

However, it is also possible that the compact conterparts of these
superluminous sources are more modest black holes, with masses in the
range of galactic black hole binaries, if mild beaming is allowed.
These possibilities will be discussed in a future paper
(King et al 2001).

Superluminous hard sources in distant young galaxies may also be an alternate
explanation for the hard deep survey sources identified with galaxies with
no sign of activitity and explained with obscured AGN (Hasinger 1999).
In the Antennae, the integrated luminosity of superluminous sources is $\sim 10^{41}
\rm ergs~s^{-1}$. Their contribution to large galaxies in formation may be substantial.

\subsection{The hot ISM}

The \chandra\ ACIS data reveal a spectacular and varied hot ISM that
had been only glimpsed at with previous ASCA and ROSAT observations (Samson et al 1996;
Read et al 1995; Fabbiano et al 1997). 
While a detailed analysis of this multiphase ISM will be
reported in a follow-up paper, here we note that  a comparison 
of the \chandra\ image with the archival HST WFPC-2 H$\alpha$-filter
image of \n4038\ (fig.~9) suggests an interesting and intricate
picture of the multi-phase ISM. The X-ray data in fig.~9 were
smoothed with a 2-pixel (1'') gaussian, to retain the small-scale
details. The X-ray contours are overlayed onto the H$\alpha$
image.  We used the reprocessed \chandra\ data with the
latest improved absolute position for this overlay (see \S2.), 
which agrees with our astrometry based on field stars detected in X-rays.
However, we have no way to check the astrometry of the H$\alpha$ image
with the available data. 

Although there is a general similarity
between the distributions of hot and cooler ISM, the two are not
identical. We find regions where H$\alpha$ and X-ray emission
are both enhanced, e.g. regions E and B, where the X-ray and H$\alpha$ peaks 
pretty much coincide, suggesting that the
H$\alpha$ and X-ray emitting gases are finely intermingled, but still
retain their different temperatures. A similar situation is observed
in the nuclear outflow of M82 (e.g. Watson et al 1984).
But we also find H$\alpha$ cavities that correspond
to peaks of the X-ray emission, suggesting  a number of 
`superbubbles', where X-ray hot gas appears to fill a hole in the H$\alpha$
emitting-gas distribution (regions A, B, possibly  C, D in fig.~9). 
Typical diameters and luminosities of these H$\alpha$-bound hot regions
are $\sim 1.5$~kpc and  $\sim 10^{39-40} \rm ergs~s^{-1}$.
These superbubbles are extraordinary, if we compare them with 
already studied superbubbles in the more nearby universe. Their X-ray
luminosities are a factor of 10-100 larger than those of the gaseous emission
of the 30 Doradus nebula in the LMC (Wang \& Helfand 1991), and of the superbubbles in
M101, that have typical luminosities of $\sim 1 \times 10^{38} \rm ergs~s^{-1}$
(Williams \& Chu 1995). Assuming temperatures of $\sim 5 \times
10^6$K and spherical emission volumes of 1.5~kpc diameter,
we derive typical masses of hot gas in these regions of $10^{5-6} M_{\odot}$.
 
The nuclear regions of the two galaxies (see \S~2.)
also appear associated with extended thermal features, as had been suggested
by Fabbiano et al. (1997). There are two extended emission
regions in the vicinities of the northern nucleus (NGC~4038, see fig.~3; region
F in fig.~9).
The southern nucleus (NGC~4039) is embedded in extended soft emission, but
it may contain an harder emission peak (figs.~3, 4). Future spectral analysis of the
ACIS data may help unravel this source. 
Note here that there
is a slight displacement between some X-ray and H$\alpha$ features (e.g., 
regions F in fig.~9, near the northern nucleus), that may be due at 
least in part to the compounded uncertainty of the HST and Chandra astrometry.
However, in other cases (e.g. region E), X-ray and H$\alpha$ peaks 
pretty much coincide.

Some of the hot X-ray emitting regions in the
ISM, and the nuclear regions themselves, may give rise to hot gaseous outflows,
like those witnessed in nearby starburst galaxies (e.g. NGC~253 and M82, Fabbiano \& 
Trinchieri 1984; Watson et al 1984; Fabbiano 1988). The luminosities of the hot gaseous regions 
in the Antennae, and therefore their thermal energy content, are comparable to or larger than 
that of the hot gaseous component of the nucleus of NGC~253.
Region B in particular shows X-ray emitting plasma surrounded by a
shell-like H$\alpha$ structure at the base of what
may be a local ouflow, as suggested by the presence of radial tendrils 
visible in a close inspection of the H$\alpha$ image.
The large-scale N-S extent of the X-ray surface brightness (\S~2.3) suggests 
the presence of large-scale hot outflowing ISM out to $\sim8$~kpc (in projection)
from the galaxy disks.

\section{Conclusions }

The \chandra\ ACIS data of the Antennae galaxies offer an excellent example of
what high (sub-arcsecond) spatial resolution, joined with spectral capabilities,
can achieve in the X-ray band. 

We have discovered a population of extremely luminous point-like sources,
with X-ray luminosities well in excess of the Eddington limit for 1~$M_{\odot}$
accretor, suggesting a young population of $\sim 10 -$(a few)$100 M_{\odot}$ black hole 
binaries, which are not seen in more evolved stellar populations, such as those 
of the Milky Way and M31. 

We detect soft thermal emission, pointing to a varied hot ISM in these
galaxies, presumably powered by the intense star formation. Features of this
hot component encompass superbubbles associated with the star-forming knots,
circumnuclear features, and an all-pervasive component, extending 
farther than the stellar bodies of the interacting galaxies and pointing to 
a galactic scale outflow.

This is our first report of these data. A detailed analysis of the
point-source population will be presented in a forthcoming paper
(Zezas et al 2001). A discussion of models of stellar and binary evolution that
may explain our data, and of their implications for the massive portion of the
IMF of spiral galaxies is the subject of another paper in preparation 
(King et al 2001). A detailed analysis of the hot ISM, and comparison with
multi-wavelength data to obtain a full picture of the multi-phase ISM,
will be given in Fabbiano et al (2001).

\acknowledgments

We thank the CXC DS and SDS teams for their efforts in reducing the data and 
developing the software used for the reduction (SDP) and analysis (CIAO).
We also thank Andrew King, Martin Ward and Martin Elvis for discussions on these
results.
This work was supported by NASA contract NAS~8--39073 (CXC)
and NAS8-38248 (HRC).

%%%%%%%%%%%%%%%%%%%%%%%%  FIGURES and FIGURE CAPTIONS  %%%%%%%%%%%%%%%%%%%%%%

\clearpage

\figcaption{Energy distribution of the integrated source counts (from the polygon 
in fig.~2, after background subtraction) before (black `x') and after (red)
removal of the hot columns (see text).}

\figcaption{Screened but otherwise unprocessed full energy range (0.1-10.0~keV)
back-illuminated ACIS-S3 image. Most of the $8' \times 8'$ S3 CCD field is
displayed. The polygon represents the source area, and the
large circles the background areas (see text). The dashed circles are regions 
excluded from the background calculation.}

\figcaption{a) Adaptively smoothed broad-band (0.1-10~keV) image of the Antennae
galaxies. b) Same as the preceeding figure, but with overlays representing the
optical countours from the POSS digital survey (white), blue crosses representing
the CO emission peaks (Stanford et al 1990), blue `X' marking the positions of the
two nuclei, and blue circles identifying three point-like sources detected with
the  ROSAT HRI (Fabbiano et al 1997).}

\figcaption{Adaptively smoothed X-ray `true' color image. Red = 1.0 - 2.0~keV;
Green = 2.0 - 5.0~keV; Blue = 5.0 - 10.0~keV.}

\figcaption{Observed ACIS-S3 spectral count distribution of the entire
emission from the Antennae (upper points; black), the coadded emission 
of detected sources (blue filled circles), 
and the diffuse emission, obtained by excluding all
the detected sources from the image (red squares). The green triangles
display the spectrum of an individual bright point-like source.
Emission lines of the hot interstellar plasma  (Fe L complex, Mg K$\alpha$, Si K$\alpha$)
are identified, as well as the 6.4~keV Fe K$\alpha$ line, visible
in the coadded point-source spectrum.}

\figcaption{The unsmoothed image of the Antennae. The polygon is the same as in
fig.~2. The circles and ellipses identify sources (or emission complexes) detected 
with {\it wavdetect}.  Dashed ellipses are complex and/or diffuse emission regions.}

\figcaption{Wide-field adaptively smoothed image, with data binned in 1" pixels.
The color scale of the X-ray image was chosen to highlight the low surface brightness
diffuse emission (in red). Peaks of the surface brightness distribution are
displayed in blue.
Optical contours are displayed on this image, which shows low-surface
brightness X-ray emission extending to the South of the optical image. 
The polygon to the South of the main body of the optical emission was used to estimate
the X-ray luminosity of this low-surface brightness X-ray component.}

\figcaption{Histogram of the luminosity distribution in NGC~4038/39 (filled - $H_o = 50$;
dashed - $H_o = 75$), compared
with that of M31.}

\figcaption{Left: X-ray image smoothed with a 1'' gaussian together with isointensity
contours. Right: X-ray contours overlayed on the HST WFPC-2 H$\alpha$ filter archival
image of the Antennae.}
\end{document}